# Enhancement of Impedance by Chromium Substitution and Correlation with DC Resistivity in Cobalt Ferrite

*Sweety Supriya,[1,a] Sunil Kumar,[1,b] and Manoranjan Kar[1,c]*

[1]Department of Physics, Indian Institute of Technology Patna, Patna-801103, India

**Abstract:** Chromium substituted cobalt ferrite with grain size less than the single domain (~ 70 nm) has been prepared by the sol-gel method. XRD analysis reveals that the samples crystallize to cubic symmetry with $Fd\bar{3}m$ spacegroup. Two transition temperatures ($T_D$~450K and $T_M$~600K) have been observed from the impedance verses temperature measurement. $T_D$ increases with the increase in frequency due to dipole response to the frequency. $T_M$ is comparable with the para-ferrimagnetic transition temperature of cobalt ferrite, which is independent of frequency. This result is well supported by the temperature dependent DC conductivity measurement. The modified Debye relaxation could be explained the impedance spectra of $CoFe_{2-x}Cr_xO_4$. The grain and grain boundary effect on impedance spectroscopy has been observed from Cole-Cole analysis. The ac conductivity follows Arrhenius behavior at different frequencies. All the samples exhibit the negative temperature coefficient of resistance behavior which reveals the semiconducting behavior of the material. The Mott VRH model could explain the DC electrical conductivity. Both ac impedance and DC resistivity are well co-related each other to explain the electron transport properties in Cr substituted cobalt ferrite. The electrical transport properties could be explained by the electron hopping between different metal ions via oxygen in the material.

**Keywords:** Cobalt Ferrite, XRD, Impedance spectroscopy, Electrical conductivity

**Introduction:**

The miniaturization of magnetic and electronic devices needs advance functional materials of nanosize to have greater efficiency in new forms. Nanocrystalline ferrites are having remarkable physical properties, which make it suitable for technological applications such as, switching devices, EMI shielding, actuators, recording tapes, microwave devices, transducers, sensors, high quality digital printing, biotechnological applications, and high density magnetic information storage.[1-12] Owing to various significant electrical and magnetic properties, ferrites have very wide applications in electronic industries. These spinel ferrites function as a dielectric material in the preferred region of frequency which ranges from radio frequency to microwave frequencies because of having a high resistivity and low eddy currents and loss factor.[7-8] Out of various ferrites, $CoFe_2O_4$ were the topic of demanding research in recent years. Cobalt Ferrite is a hard ferrimagnetic material with a Curie temperature at around 793 K has distinctive properties like high magnetic sensitivity (5400 Oe), high magnetocrystalline anisotropy and moderate saturation magnetization (80emu/g).[5, 9-11]

Now days there have been an intensive demand for dielectric and magneto-dielectric application based materials.[12] Ferrite materials are insulating magnetic oxides which permit electromagnetic field penetration as compared to metals, where there is limitation of penetration of the high frequency field due to skin effect.[13] Partial replacement of Fe by metal ions in $CoFe_2O_4$ proposed to tailor the structural, magnetic, electrical and dielectric properties.[14-19] This replacement causes structural distortion and induces lattice strain and modifies the magnetic and electrical properties significantly.[16-19] There are abundance researches on cobalt ferrite for its application as a magnetic material. Researchers have modified its magnetic property by substituting either in Co site and/or in Fe site. However, there is limited study of its dielectric properties to explore the technological applications. In this regard, there are a few reports available on dielectric properties of cobalt ferrite.[16-22]

Cobalt ferrite exhibits an inverse spinel structure represented by the formula $AB_2O_4$ as $CoFe_2O_4$. In which metal cations $Fe^{3+}$ and $Co^{2+}$ are at A (tetrahedral) and B (octahedral site) interstitial sites respectively formed by the face centered cubic (FCC) arrangement of oxygen anions.[23] Cobalt ferrite crystallizes to normal/inverse/mixed phase spinel ferrite. However, mostly the inverse spinel structure is obtained by the conventional method of preparation. In inverse spinel FCC lattice, half of $Fe^{3+}$ ions are occupied by tetrahedral (A) sites of spinel structure and half of $Fe^{3+}$ and $Co^{2+}$ ions are occupied by the octahedral (B) sites. The distance between the two metal ions one on B site and another at A site is larger than the distance between metal ions in the same site. The electrical conduction mechanism in ferrites is mainly explained by considering the polaron hopping between $Fe^{3+}$ to $Fe^{2+}$ ions.[24] Mostly it arises due to local displacement of charge and results in polarization. The probability of electron hopping between B–B hopping is very high compared to that of B-A hopping. There are only $Fe^{3+}$ ions at A sites for inverse spinel structure, hence there is no possibility of hopping between A−A sites. Chromium ions have a preference for octahedral site, thus they reduce the $Fe^{3+}$ ions in the Fe-site and in turn reducing $Fe^{3+}$ - $Fe^{2+}$ ions polaron hopping.[25] Also, this concept has been extended to understand the dipole oscillation in spinel cobalt ferrite. It is observed that, the charge hopes from $Fe^{2+}$ to $Fe^{3+}$ or vice-versa via oxygen. Also, there is possibility of charge hopping in between $Co^{2+/3+}$-O-$Fe^{2+/3+}$ or $Co^{2+}$-O-$Co^{3+}$. Due to the above assumptions, a dipole form between $Co^{2+/3+}$/$Fe^{2+/3+}$ and it oscillates.

The electrical resistivity in the ferrites significantly depends on the density, porosity, grain size, chemical composition and crystal structure of the sample.[26-27] Although the high dielectric loss of cobalt ferrite at low frequency confines their electrical applications, the dielectric relaxation behavior of nanocrystalline cobalt ferrite which is reported recently, indicates its potential application in phase shifters.[28] Moreover, the successive observation of relaxation behavior in magnetic materials may break through their limitation in magnetic applications and extends widely to electrical aspect. Also, the factors involved in tailoring the magnetic and electrical properties of polycrystalline materials are method of preparation, temperature, sintering time, cation distribution, type and quantity of dopants and grain size.[26-27] Mostly the surface characteristic strain at lattice site and charge distribution in octahedral and tetrahedral sites modifies to tune the physical properties. Decrease in resistivity is observed with the increase in temperature only because of the decrease in grain boundary resistance. The electrical properties reveal the behavior of localized electric charge carriers and the phenomenon correlated to the polarization. The macroscopic electrical properties of polycrystalline ceramics can be well explained by the microstructure grain and grain boundary distribution in the sample. These factors stimulate a deep study of frequency and temperature dependent impedance spectroscopy through experimental and theoretical modeling. As discussed above, the electrical properties of $CoFe_2O_4$ can be controlled by suitable substitution at Co or Fe site as it will modify the charge distribution as well as lattice strain. Cr is a transition element with similar charge states of Fe with different ionic size. Hence, there are a few reports on Cr substitution on $CoFe_2O_4$ to tune the magnetic properties. It is observed that the hard ferrimagnetic cobalt ferrite becomes a soft magnetic material by Cr substitution.[29] Hence, it is expected there will be a modification of electrical properties on $CoFe_{2-x}Cr_xO_3$. However, a literature survey shows that, there is no detail report available to understand the electrical transport properties of Cr substituted cobalt ferrite. Moreover, researchers have reported dielectric properties of different substituted cobalt ferrite or DC transport properties independently. Hence, the study of correlation between ac impedance spectroscopy and DC resistivity is necessary to understand the electrical transport properties on substituted cobalt ferrite. Also, it is expected that the magnetic ordering in the cobalt ferrite affects the electrical transport properties. To address the above issues the Cr substituted cobalt ferrite has been prepared. The detailed temperature variation impedance spectroscopy and DC resistivity have been studied. Interestingly, it is observed that the magnetic ordering temperature well reflected in both DC resistivity and ac impedance measurement. The magnetic ordering temperature is independent of applied ac signal frequency. The present study can help to understand the other substituted cobalt ferrite and opens a window to tune the electrical properties of cobalt ferrite by substitution for technological applications.

**Experiments:**

Nanocrystalline samples of $CoFe_{2-x}Cr_xO_4$ (CFCO) ferrite for x=0.0, 0.1, 0.2, 0.3 and 0.4) were prepared by the modified sol-gel technique.[30-33] The precursor materials are Cobalt nitrate ($Co(NO_3)_2 \cdot 6H_2O$) (Merck), Iron nitrate ($Fe(NO_3)_3 \cdot 9H_2O$)(Alfa Aesar), Chromium nitrate ($Cr(NO_3)_3 \cdot 9H_2O$)(Merck) and Citric acid ($C_6H_8O_7 \cdot H_2O$)(Merck) with 99.9% purity. The powders of the pre-sintered synthesized material were ground in a mortar pestle and heated at 1173 K for 1 hour with the heating rate of 5 K per minute. The samples were made into pellets having diameter 10 mm and the cross-sectional area is 7.85 cm$^2$. Phase identification and crystal structure of the materials were investigated by using the X-ray diffraction (XRD) technique with the help of Rigaku X-ray Diffractometer (Model TTRX III). Measurements were carried out at room temperature using CuK$\alpha$ radiation ($\lambda$=1.5406Å). The density of the material measured by employing the Archimedes' principle and with the help of a balance Sartorious (Model No. CPA225D). Hardness of the materials was measured by the Micro Hardness Tester (Model No. MICROTEST MTR3/50-50/NI) using Vicker indentors under a test force 2 N for 5 second. Electrical measurements were carried out on the specially prepared samples. Briefly, the surfaces of the samples were well polished, silver paint was coated on the surface and then heated in a furnace programmed as at 150 °C for an hour and 250 °C for half an hour to remove the content of binder (PVA-Polyvinyl alcohol, hot water soluble form HIMEDIA laboratories Pvt. Ltd.) used for making pellets and removal of epoxy used for making electrodes. Pellets with silver electrode were sandwiched between two metal holders and connected through wires to N4L Impedance analysis interface (PSM1735 NumetriQ). The sample assembly was kept in a furnace equipped with a temperature controller to carry out the impedance measurements as a function of frequency and temperature. The sample assembly constitutes the parallel plate capacitor geometry with the $CoFe_{2-x}Cr_xO_4$ as the dielectric. An ac signal of 2V and frequency (f) in the range of 1 Hz to ~10 MHz was applied to the circuit using N4L Impedance analysis interface through PSM1735 NumetriQ. In addition, standard calibration was carried out before the measurement to avoid any stray capacitance, contact and lead resistance and, reset the device to remove the history of data. The capacitance, dielectric, phase, loss tangent and impedance are recorded with varying temperatures from 323 K to 673 K as a function of frequency ranging from 1 Hz–10 MHz to quest the dielectric dispersion and the response of Cr substituted CFO in the conduction mechanism. The Complex dielectric, complex electric modulus and the conductivity were calculated at different temperatures using measured capacitance and phase data at the fixed frequency. The temperature variation of DC resistivity measurement was carried out by Keithley multimeter (2001) using two probe method.

**Results and Discussion:**

X-ray diffraction (XRD) patterns of $CoFe_{2-x}Cr_xO_4$ for x=0.0, 0.1, 0.2, 0.3 and 0.4 are shown in the figure 1(a). The observed peaks in the diffraction patterns in the XRD pattern could be indexed to $Fd\bar{3}m$ space group with cubic crystal symmetry. The extra peaks and intermediate phase formation were not observed within the X-ray diffraction pattern technique limit. The observed broadening of the diffraction peaks in the XRD pattern affirms that the particles are of ultrafine nature and small in size. The crystallite size of compound $CoFe_{2-x}Cr_xO_4$ for x=0. 0, 0.1, 0.2, 0.3 and 0.4 was calculated using Debye-Scherrer's formula.[34] Average crystallite sizes are found to be in between 60-70 nm which are enlisted in table I. Hence particles are below the single domain size as the single magnetic domain size of $CoFe_2O_4$ is ~ 70 nm .[35] The (311) peak, which is highest intensity peak in $CoFe_2O_4$ XRD pattern is shown in figure 1(b) for all the samples. It is observed that peak shifts towards higher angle. It reveals the decrease of lattice parameter enlisted in table 1 with the increase in Cr substitution. It is due to the smaller ionic radius of Cr to that of Fe and it also reveals the incorporation of Cr in the $CoFe_2O_4$ lattice. The lattice parameters are comparable with those reported in the literature for cobalt ferrite.[36] As the density is very important for impedance study it was determined by the Archimedes' method and those values are enlisted in table I. These values are comparable to those reported for cobalt ferrite i.e., 5.3g/cm$^{-3}$ (approx) of theoretical value.[36] Also values are almost constant for all the samples (~4.6 g/cm$^3$), and above 86% of the theoretical value(5.3 g/cm$^3$). [37] Hence the variation of electrical properties is only due to Cr substitution as samples are well dense, which is explained in the next section. Also the hardness of the samples was measured and those values are enlisted in the table I. It is also observed that hardness for all the samples are close to each other (154-160 HV) and justified with the earlier reports.[38]

**Impedance Analysis:**

Impedance spectroscopy is an important technique to characterize the electrical properties of the material and their electrode interface.[39] Impedance spectroscopy techniques employed to study the electric properties of $CoFe_{2-x}Cr_xO_4$ for x=0.0, 0.1, 0.2, 0.3 and 0.4. The contribution from the intergrain and intragrain effect due to heterogeneous distribution of charge and their influence on electrical conductivity as a function of frequency and temperature has been studied extensively.

This technique also divides the resistive and reactive part of electrical component that gives the exact picture of electrical properties. The temperature and frequency dependent real ($Z'$) and imaginary parts ($Z''$) of the Complex Impedance $Z^*$ are [40-41] described as,

$$Z^* = Z' - jZ'' \qquad (1)$$

$$Z' = \frac{R_p}{\left[1 + (\omega R_p C_p)^2\right]} \qquad (2)$$

$$Z'' = \frac{\omega R_p^2 C_p}{\left[1 + (\omega R_p C_p)^2\right]} \qquad (3)$$

Where $R_p$ is the parallel resistance, $C_p$ is the parallel capacitance, $\omega$ is the angular frequency.

Mostly these behaviors explained by the Koop's theory, which describes that the inhomogeneous dielectric structure consists of fine conducting grains and poorly conducting grain boundaries.[42] Hence the conductivity and impedance of grains are high from that of the grain boundary. Figure 2(a) - 2(e) shows the frequency dispersion of real parts of the impedance ($Z'$) of $CoFe_{2-x}Cr_xO_4$ for x= 0.0, 0.1, 0.2, 0.3, and 0.4 samples at some selected temperatures. The logarithmic scale of frequency is taken for clarity. The frequency variation of $Z'$ from temperature 473 K to 673 K are shown as insets in the figures 2(a) - 2(e). It is observed that the magnitude of the real part of impedance ($Z'$) decreases with the frequency, which is the normal behavior of ferrites[43] and follows Koop's theory as discussed above. There is a sharp decrease in $Z'$ at the lower frequencies and eventually decreases towards high frequency and remain constant at very high frequencies which indicate the dispersion of $Z'$ at low frequency. This type of behavior observed due to the polarization of valence states of cation and space charge polarization. Impedance is found to be non-dispersive at high frequency range (1 MHz-10 MHz). This happens because the dipolar polarization and surface charge polarization fail to flip with respect to the fast changing of applied alternating electric field due to high frequency. Basically the impedance at very low frequency is due to contribution from space charge polarization, dipolar polarization, atomic polarization and electronic polarization. As frequency increases, polarization due to dipolar and space charge abruptly decreases and approach to zero at high frequency. At higher temperature, charge carrier gets momentum, due to high thermal energy, so the impedance dispersion is observed at high frequency too. The signature of transition from high impedance to low impedance value at a particular frequency as shown in figure 2 remains unaltered with the increasing temperature. Hence this signature could be leads to use the material frequency filter or frequency band selector at various temperatures. The electric dipole polarization in the cobalt ferrite can be understood by considering the electronic interaction between cations.[44] The electronic exchange between $Fe^{2+}$ and $Fe^{3+}$ via oxygen causes the confined displacement of electrons along the direction of electric field, which determines the polarization and relative impedance behavior. As the resistance of grain boundary is very high, then the electrons pile up at the interface which is the cause of surface charge polarization. Hence, the applied voltage drop in the specimen could be mainly arises across grain and grain boundary, causing space charge accumulation under the influence of electric field.[45]

The magnitude of impedance is the order of mega ohm which is matched well with the earlier reports.[46] It has been observed that (figure 2) $Z'$ increases with the increase in Cr doping concentration. Generally the increase in impedance can be attributed to the grain size distribution in the oxides. However, the grain as well as crystallite distribution in the present samples is very narrow. Hence, the increase in $Z'$ with the increase in Cr is due to the substitution effect. The ionic size of Cr is smaller compared to that and Fe which creates lattice strain and hence modification of dipole strength. The details of temperature dependence $Z'$ are discussed in the later section.

Figures 3(a)-3(e) show frequency variation of imaginary (i.e. ac loss) part of impedance ($Z''$) as a function of temperature. The logarithmic scale of frequency is taken in each graph in order to have clarity of presentation. Each figure has an inset which depicts the frequency variation of $Z''$ from temperature 473 to 673 K. A peak has been observed for all the samples in frequency variation of $Z''$ spectrum, which can be seen from the figures 3(a)-3(e). Peaks are very broad and asymmetric in nature. The appearance of the peak represents the presence of electrical relaxation and the distribution of the relaxation times, which attributed to the asymmetrical broadening of the peaks. The maximum loss on the $f_{max}$ position shifts to a higher frequency with the increase of temperature. It reveals that maximum loss part is temperature dependent. The amplitude of $Z''$ maximum peak decreases with increase in temperature. The magnitude of the $Z''$ at $f_{max}$ peak frequency increases with the increase in Cr concentration for a particular temperature, it is because of the presence of inhomogeneous atoms at lattice site. The value of a frequency corresponding to $Z''$ maximum peak at a certain temperature is shown in table II. The frequency $2\pi f_{max}$ related to $Z''_{max}$ could be modeled through the Arrhenius law as given in equation 4 where $\omega_o$ is the pre-exponential factor and $E_A$ is the activation energy. The activation energy was calculated from the slope of the plot $ln(\omega_{max})$ verses $1000/T$ and given in table V.

In this section, the temperature dependence of impedance study is discussed. Figures 4 (a)-4(e) shows the temperature dependence of real part ($Z'$) of impedance as a function of frequency. Figures 4(a)-4(e) comprises of three categories, i.e., i) Figures 4(a1)-4(e1) are shown for the frequencies 1097 Hz and 10722 Hz, ii) Figures 4(a2)-4(e2) are shown for the different 6 frequencies in the range of 104761 to 3199267 Hz and iii) insets show enlarged version of the $Z'$ variation in the range of 500 to 873 K to represent the clarity in peaks. It is observed that, the magnitude of $Z'$ increases with increase in Cr concentration. The similar trend has been observed by other research groups in substituted cobalt ferrite.[24] An increase in Cr concentration in $CoFe_{2-x}Cr_XO_4$, causes an increase in the impedance ($Z'$) because of strain introduced by inhomogeneous atoms at lattice sites, as discussed in earlier sections. There are two peaks have been observed at different temperatures. However, earlier reports on $CoFe_2O_4$ exhibits only single peak. It is worth noting that earlier report for single peak [47] were for micrometer grain size samples, however the present samples are well dense nanocrystallites. Hence this interesting observed feature has been studied in detail.

It is difficult to find the exact peak position of the curve (figure 4) as the transition is very broad. Hence the peak temperature was determined from the $\left|\frac{dZ'}{dt}\right| \sim T$ plot (figure not shown). The maxima were noted at $\left|\frac{dZ'}{dt}\right| \to 0$. The 1$^{st}$ peak transition temperature around 450 K ($450 \pm 100K$) is named as $T_D$ and $T_M$ is called to the 2$^{nd}$ peak (2$^{nd}$ transition temperature) around 600 K ($600 \pm 20K$). Both the transition temperatures $T_D$ and $T_M$ are enlisted in the table III. ($T_D$) for each composition shifted towards higher temperature with the increase in frequency. This signature is consistent with earlier reports[24] which is the typical impedance behavior of the cobalt ferrite. This is because at higher frequency, the dipoles lag behind the fast changing alternating applied electric field. At higher frequency, polarization decreases and hence the impedance. But temperature is also increasing with increasing frequency, henceforth to maintain the polarization with rapid change in frequency, increasing temperature play a very crucial role in order to contribute energy to maintain this polarization. Eventually temperature in the form of energy supplied to maintain polarization, and this result in the form of peak shift of $Z'$ with the increase in frequency and temperature.

$T_D$ shifts more than 200 K by changing the frequency from 1 KHz to ~10 MHz, however, the second peak position ($T_M$) shifts <20 K, which is clear from figure 4 and enlisted in the table III. It reveals that the second peak almost does not respond to the frequency. is very less. Paramagnetic to ferrimagnetic transition temperature for ~ 70 nm cobalt ferrite sample is around 600 K[48-49] which is close to the second observed peak in temperature verses $Z'$ plot for the present work. The magnetic ordering temperature ($T_M$) is lower for nanocrystalline substituted cobalt ferrite compared to that of bulk substituted cobalt ferrite.[50] It is independent of applied electric field frequency because the electric field does not respond to magnetic dipole in the cobalt ferrites and hence a weak magnetoelectric coupling material, which has been almost established. The impedance at $T_M$ is changed due to only change in electrical resistive part of the sample which arises due to the presence of magnon. The resistive part of impedance is frequency independent. Hence, the transition temperature $T_M$ is independent of frequency. Earlier reports on the impedance of cobalt ferrites are mostly in bulk sample where the magnetic transition temperature is very high. Hence it was not feasible to observe the para-ferrimagnetic transition temperature. Although there are a few reports on nanocrystalline sample, but the impedance measurement has been reported for a limited temperature i.e. <500K. However, in the present report the magnetic transition is clearly observed. Hence the present study opens the window to understand the different magnetic and electrical properties of cobalt ferrite. One can easily evidence the above characteristic by DC measurement where the second transition temperature ($T_M$) will prominently compared to the 1st peak due to electron magnon scattering, and discussed in the later section.

The magnitude of Z' decreases with the temperature above $T_M$. This behavior accounts the semiconducting nature of the sample at high temperature. That is a negative temperature coefficient of resistance (NTCR). This semiconducting behavior with increasing temperature leads to hopping of charge carrier and results in increase of ac conductivity which is discussed in the later section.

The plots of temperature dispersion of the imaginary parts of impedance Z'' of $CoFe_{2-x}Cr_xO_4$ for x=0.0. 0.1, 0.2, 0.3 and 0.4 are shown in the figures 5(a)-5(e). Insets in figures 5(a1)-5(e1) show the temperature dependent Z'' spectrum of $CoFe_{2-x}Cr_xO_4$ for x= 0.0. 0.1, 0.2, 0.3 and 0.4 in enlarge form to identify the transition temperature $T_M$. The magnitude of Z'' is high at low temperature and gradually decreases towards high temperature and eventually Z'' became independent of temperature. The magnitude of Z'' is decreasing with increasing frequency (because dipole lag behind increasing frequency) and temperature (because of semiconducting behavior).

When an alternating electric field is applied the dipoles oscillates in its mean position. This oscillation can be modeled to one of the electric oscillators such as the parallel RC or LCR oscillator. A peak of maxima is observed when the frequency of the dipole oscillation is equal to frequency of applied electric field, called the resonant frequency. The condition for the resonance is $\omega\tau=1$, where time constant $\tau=RC$. The amplitude of the oscillations will be small for any frequency other than the resonant frequency. The electrical charge transport through grain, grain boundary, partially electrode interface modeled through consecutive parallel RC circuit. The resonant frequency stands as $\omega_r =1/RC$.

The grain, grain boundary and electrode interface contribution can be observed individually from total impedance spectra .[39] Fig 6(a)-6(d) shows the Cole-Cole plots or complex impedance spectra for samples $CoFe_{2-x}Cr_xO_4$ for x=0.0. 0.1, 0.2 and 0.3 at selected temperatures. The Cole-Cole plot gives the complete involvement of grain and grain boundaries. The plot exhibits single semicircle which indicates that the contribution of electrical conductivity mainly arises from the grain, grain boundary and electrode interface. The semicircle obtained at low frequency corresponds to grain boundary resistance while semicircle at high frequency corresponds to grain resistance.[39] As it has been observed that there are two kinds of semicircles, one in low frequency applied electric fields illustrates the grain boundary resistance and another in high frequency applied electric field region indicates the role of grain resistance. The equivalent circuits based on impedance data are shown in the figure 7. The parameter $R_s$, $R_g$, $C_g$, $R_{gb}$, $C_{gb}$, *CPE and n* correspond to series resistance, grain resistance, grain capacitance, grain boundary resistance, grain boundary capacitance, constant phase element and exponential power *n*. All the parameter values are obtained by modeling the impedance data to an appropriate electrical circuit oscillator. The electrical circuit models were fitted using the ZSimpWin software in which frequency is implicit in every curve. The values of electrical elements obtained from the analysis are enlisted in the table IV.

From the Cole-Cole plot, the semicircular arc depresses as the temperature increases and the intercepts of arc with the *Z'* axis shifted towards the origin with the increase of temperature. The depression of an arc indicates that the dielectric relaxation deviates from the ideal Debye behavior and the shifting of arc intercept towards the origin pointed that there is a decrease in bulk resistances. As we observed from the table IV, that $R_g$ and $R_{gb}$ decrease with the increase in temperature, which once again validates the insulator or NTCR of semiconductor. It is also perceived that the grain boundary resistance increases with the increase in Cr concentration in the cobalt ferrite. It implies that the Cr substitution enhances the barrier property not due to its influence against the flow of charge carriers.

**AC Conductivity Analysis:**

Plots of variation of ac conductivity with temperature of compound $CoFe_{2-x}Cr_xO_4$ for x=0.0, 0.1, 0.2. 0.3 and 0.4 are shown in the figures 8(a)-8(e). The temperature dependence of ac conductivity can be analyzed by employing the Arrhenius equation[45]

$$\sigma_{ac} = \sigma_o e^{\frac{-E_A}{K_B T}} \quad \text{or} \quad \omega_{max} = \omega_o \exp{\frac{-E_A}{kT}} \tag{4}$$

Where $\sigma_o$ is the electrical conductivity at infinite temperature. $E_A$ is the activation energy and $K_B$ is the Boltzmann constant. The activation energy is calculated at different frequencies of the compounds $CoFe_{2-x}Cr_xO_4$ for x=0.0, 0.1, 0.2, 0.3 and 0.4. The value of $E_A$ obtained at 533669 Hz for all the compositions are enlisted in the table V. This indicates that the ion overcomes the different barriers in the respective regions of conductivity. It is observed from the figure 8(a)-8(e) that the curve comprises of three parts. In region I, below 423 K conductivity was frequency dependent. The increase in frequency causes an increase in hopping of charge carriers. Therefore, increase in ac conductivity has been observed with the increasing frequency. It is concluded that strong frequency dispersion exists in the region I (313 K to 423 K) and $\sigma_{ac}$ increases with increase in temperature. The strong dispersion caused by polarization due to hopping of charge carriers at random sites of different barrier heights and separation. A decreasing trend of activation energy has been noticed by the increase in Cr concentration. In region II (423 K to 543 K), conductivity depends on both temperature and frequency. And conductivity in this region is a contribution from short range oxygen vacancies. Followed to region III (543 K to 673 K), electrical conductivity is independent of frequency and only depends on temperature. The electrical conductivity in this range plays a role due to long range vacancy and the creation of a defect.[51] Here the increasing fashion of activation energy was studied with the increase in Cr concentration. It indicates that relatively more activation energy is required in the form of thermal energy for the hopping of charge carriers due to having heterogeneous atoms at lattice site. The increase in electrical conductivity with increase in temperature at all the frequencies was observed. It resembles the semiconducting behavior which represents a negative temperature coefficient of resistance (NTCR). The figure 8 represents the Arrhenius plot of the ac conductivity at different frequencies for $CoFe_{2-x}Cr_xO_4$ with x=0.0, 0.1, 0.2, 0.3 and 0.4 compounds.

The figures 9(c) – (e) depict the frequency dependence of ac conductivity of $CoFe_{2-x}Cr_xO_4$ compounds for x=0.0, 0.1, 0.2, 0.3 and 0.4 at different temperatures. The frequency dependence of ac conductivity curves shows the response of the material to the applied time varying field. The nature of transport process in the material will be investigated through these studies. The ac electrical conductivity was determined by the equation;[41]

$$\sigma_{ac} = \varepsilon_o \varepsilon' \omega \tan(\delta) \qquad (5)$$

At low frequency, more charge accumulation takes place at the electrode interface region due to space charge polarization, called polarization region, which acts as a barrier in the conduction, so conductivity is low in this region. In the middle region, the conductivity is almost independent of frequency called plateau region and equal to DC conductivity which is associated with drifting of charge carriers. At high frequency, the conductivity increases with frequency, called dispersion region.[52] In this region frequency is very high, so space charges get dispersed rather than relaxation, caused by localized charge carriers. The frequency region at which change of slope takes place is called hopping frequency. The hopping frequency increases with the increase in temperature.

The behavior of the curve is showing an increase in ac conductivity with frequency. The curve consists of two parts: DC conductivity part, characterize by plateau behavior seen at low frequencies. The second part is ac conductivity which increases at high frequencies. AC conductivity increase with the increase in the frequency for all the compositions. There is a gradual rise in conductivity at low frequencies, whereas conductivity increases sharply at high frequencies (figure 9). The cause of this conduction is grain boundary effect which act like a hindrance to the mobility of charge carriers. The presence of ionic part of the grain boundary is the cause of conduction at high frequency. And this linear increase in AC conductivity with the increase in frequency confirms the polaron type of conduction. It is consistent with the earlier report.[53] The DC resistivity has been studied to understand the polaron hopping mechanism which is discussed in the next section.

**DC Resistivity Analysis:**

The electrical behavior of $CoFe_{2-x}Cr_xO_4$ compounds for x=0.0, 0.1, 0.2, 0.3, and 0.4 are analyzed via measuring the DC resistivity. Temperature variation of resistivity is shown in figures 10(a)-10(e). It has been observed that resistivity decreases with increase in temperature and validates the behavior of semiconductor as negative temperature coefficient of resistance. It is observed that there are two anomalies in the temperature dependent DC resistivity curve. Transition temperatures, $T_D$ and $T_M$ were obtained from the temperature dependent DC resistivity measurement and these two temperatures are consistent with the peaks observed in the $Z'$ vs. temperature plot which was discussed in the previous section (Figure 4 & table III).

Chromium substitution in $CoFe_2O_4$ generally causes disorder in the system which may tend to localize the charge carriers at the doping site. The electrical resistivity data were analyzed by employing small polaron hoping (SPH) and Mott's 3-d variable range hoping model.[54, 55] However, the lowest reliability factor was obtained from the analysis by employing Mott's variable range hopping (VRH) model which is given as;

$$\rho(T) = \rho_o \exp(\frac{T_o}{T})^{\frac{1}{4}} \qquad (6)$$

Here $\rho_o$ is the residue resistivity $T_o$ is the Mott Characteristic temperature. The DC resistivity verses temperature plots are shown in figures 14(a-e) along with the theoretical curve generated by the Mott VRH model. It is observed that in the whole temperature range, there are four regions. However, only two regions could be analyzed by the VRH model. The resistivity at room temperature ($\rho_{300K}$) and, calculated residual resistivity ($\rho_o$) and characteristic temperature ($T_o$) are enlisted in table VI. All the parameters ($\rho_{300K}$, $\rho_o$ and $T_0$) increases with the increase in Cr concentration, which is consisted with the impedance studies. Hence it is concluded that the transport properties of Cr doped Cobalt ferrite could understood by the VRH model. Also, it is worth noting that the hopping difference and hopping energy is different at two regions.

**Conclusion:**

Chromium substituted nanocrystalline cobalt ferrite has been synthesized by the citrate precursor method. Samples are crystallized to cubic crystal symmetry (spacegroup $Fd\overline{3}m$). The temperature variation of impedance spectroscopy reveals the two transitions in the material. First transition temperature $(T_D)$ is observed at around 400K, which increases with the increase in the frequency. Second transition temperature $(T_M)$ at around ~600K is identified as the para-ferrimagnetic transition temperature. Both the transition temperatures are also identified from the precise DC resistivity measurement. The above results open a window to understand the electrical behavior of cobalt ferrite. The conduction mechanism of the chromium substituted cobalt ferrite followed the VRH model. The hopping distance below and above $T_D$ is different. The impedance and DC resistivity on the material could be explained by the hopping mechanism between $Fe^{3+}/Co^{3+}/Cr^{3+}$ and $Fe^{2+}/Co^{2+}/Cr^{2+}$ via oxygen.

The ac conductivity with temperature variation shows Arrhenius behavior. The conductivity is found to decrease with the increase chromium substitution. This result also supports the observation of an increase in $R_{gb}$ with the increase in Cr concentration. $R_{gb}$ was obtained by modeling electrical equivalent circuit via Cole-Cole plot. The depression of an arc has been observed in the Cole-Cole plot which shows the deviation of dielectric relaxation from the ideal Debye behavior and the shifting of arc intercept towards the origin shows the decrease in bulk resistances and validates NTCR properties of semiconductors. The activation energy ($E_A$) derived from the Arrhenius behavior obtained from ac conductivity and $ln(\omega_{max})$ verses $1000/T$ using $Z''_{max}$ is found to be different because of different kinds of barrier in against the flow of charge carriers. This study opens a window to understand the electrical transport properties in cobalt ferrite and its modification by substitution for technological applications.

**Acknowledgement**

Authors are thankful to Council of Scientitific Industrial Research, Department Of Science & Technology and Department of Atomic Energy, India vide sanction number 03/1183/10/EMR-II, SR/FTP/PS-103/2009 and 2011/20/37P/03/BRNS/007 respectively for financial assistance and also UGC-ref. No.: 4050/ (NET-JUNE 2013) for JRF. The authors also acknowledge IIT Patna for providing the working platform. Authors are thankful to Dr. Anup Keshri, Department of Materials Sceince Engineering, IIT Patna for his help for hardness testing.

Table I. Crystallite size and lattice parameter, density and Vicker's hardness of $CoFe_{2-x}Cr_xO_4$ for x= 0, 0.1, 0.2, 0.3 & 0.4 samples.

| Sample | Crystallite size (nm) | $a = b = c$ (Å) | Density ($g/cm^3$) | Vicker's Hardness (HV) |
|---|---|---|---|---|
| x=0.0 | 65 ± 1 | 8.377(5) | 4.60 | 160 |
| x=0.1 | 63 ± 1 | 8.376(2) | 4.65 | 156 |
| x=0.2 | 68 ± 1 | 8.373(1) | 4.63 | 159 |
| x=0.3 | 68 ± 1 | 8.373(9) | 4.60 | 154 |
| x=0.4 | 65 ± 1 | 8.376(2) | 4.62 | 156 |

Table II. Peak frequency obtained from Z" vs. frequency plot of $CoFe_{2-x}Cr_xO_4$ for x= 0, 0.1, 0.2, 0.3 & 0.4 samples.

| Temperature (K) | Frequency (Hz) | | | | |
|---|---|---|---|---|---|
| | x=0.0 | x=0.1 | x=0.2 | x=0.3 | x=0.4 |
| 323 | 576 | 568 | 112 | 572 | 95 |
| 373 | 9179 | 9101 | 3430 | 792 | 932 |
| 423 | 88831 | 75713 | 33516 | 6579 | 7742 |
| 473 | 632463 | 535391 | 200923 | 46415 | 64280 |
| 523 | 2738875 | 1215646 | 739072 | 278255 | 453487 |
| 573 | 3790914 | 2321528 | 1417470 | 5336]69 | 200923 |
| 623 | 3778943 | 1965804 | 739072 | 533669 | 236448 |
| 673 | 5213515 | 7753431 | 1963040 | 1023530 | 739072 |

Table III. Transition temperatures obtained from impedance vs. temperature measurement of $CoFe_{2-x}Cr_xO_4$ for x= 0, 0.1, 0.2, 0.3 & 0.4 samples.

| Frequency(Hz) | *Transition temperatures (K)* | | | | | | | | | |
|---|---|---|---|---|---|---|---|---|---|---|
| | x=0.0 | | x=0.1 | | x=0.2 | | x=0.3 | | x=0.4 | |
| | $T_D$ | $T_M$ | $T_D$ | $T_M$ | $T_D$ | $T_M$ | $T_D$ | $T_M$ | $T_D$ | $T_M$ |
| 1097 | 343 | 592 | 343 | 585 | 343 | 603 | 373 | 593 | 311 | 584 |
| 10722 | 383 | 592 | 383 | 604 | 412 | 632 | 433 | 593 | 333 | 595 |
| 104761 | 433 | 594 | 433 | 604 | 453 | 613 | 493 | 603 | 483 | 573 |
| 327454 | 473 | 604 | 463 | 604 | 493 | 613 | 553 | 603 | 553 | 603 |
| 533669 | 473 | 613 | 473 | 602 | 503 | 613 | 563 | 613 | 523 | 603 |
| 739072 | 493 | 613 | 483 | 604 | 523 | 603 | 573 | 613 | 533 | 613 |
| 1023531 | 553 | 623 | 503 | 595 | 533 | 623 | 573 | 623 | 523 | 623 |
| 3199267 | 543 | 623 | 553 | 615 | 553 | 623 | 583 | 623 | 593 | 623 |

Table IV. Parameters of series resistance($R_s$), grain resistance($R_g$), grain capacitance ($C_g$), grain boundary resistance ($R_{gb}$), grain boundary capacitance ($C_{gb}$), constant phase element (*CPE*) and *n* obtained from the Cole-Cole plot and analysis of $CoFe_{2-x}Cr_xO_4$ for x= 0, 0.1, 0.2, 0.3 & 0.4 samples.

| | T(K) | Rs(Ω) | $C_g$(F) | $R_g$(Ω) | $C_{gb}$(F) | $R_{gb}$(Ω) | CPE | n |
|---|---|---|---|---|---|---|---|---|
| $CoFe_2O_4$ | 323 | 4.615 | $1.16 \times 10^{-9}$ | $6.537 \times 10^5$ | $3.39 \times 10^{-10}$ | $9.354 \times 10^5$ | $4.46 \times 10^{-8}$ | 0.37 |
| | 373 | 5.42 | $1.06 \times 10^{-9}$ | $5.206 \times 10^4$ | $3.60 \times 10^{-10}$ | 56610 | $2.91 \times 10^{-7}$ | 0.3271 |
| | 423 | 6.817 | $4.99 \times 10^{-10}$ | 2119 | $6.45 \times 10^{-10}$ | 7977 | $9.87 \times 10^{-8}$ | 0.4163 |
| | 473 | 5.345 | $3.69 \times 10^{-10}$ | 489.1 | $7.74 \times 10^{-10}$ | 1288 | $2.41 \times 10^{-7}$ | 0.01614 |
| | 523 | 4.036 | $2.96 \times 10^{-10}$ | 218.9 | $5.46 \times 10^{-9}$ | 277.7 | $6.58 \times 10^{-9}$ | 0.0964 |
| | 573 | 9.0776 | $3.40 \times 10^{-10}$ | 169.4 | $1.57 \times 10^{-9}$ | 196.5 | $6.14 \times 10^{-9}$ | 0.0164 |
| $CoFe_{1.9}Cr_{0.1}O_4$ | 323 | 5.817 | $8.47 \times 10^{-10}$ | $9.284 \times 10^5$ | $2.72 \times 10^{-10}$ | $9.88 \times 10^5$ | $2.27 \times 10^{-8}$ | 0.4039 |
| | 373 | 9.946 | $1.16 \times 10^{-9}$ | $3.993 \times 10^4$ | $7.02 \times 10^{-10}$ | $7.02 \times 10^4$ | $2.068 \times 10^{-7}$ | 0.3216 |
| | 423 | 11.94 | $4.09 \times 10^{-10}$ | 3754 | $7.13 \times 10^{-10}$ | 7860 | $2.108 \times 10^{-7}$ | 0.3874 |
| | 473 | 13.05 | $3.38 \times 10^{-10}$ | 939.4 | $1.57 \times 10^{-9}$ | 4230 | $2.20 \times 10^{-7}$ | 0.01234 |
| | 523 | 13.4 | $9.90 \times 10^{-11}$ | 753.7 | $2.76 \times 10^{-10}$ | 801.7 | $2.29 \times 10^{-9}$ | 0.1258 |
| | 573 | 18.76 | $3.29 \times 10^{-10}$ | 238.3 | $5.99 \times 10^{-10}$ | $8.95 \times 10^8$ | $4.50 \times 10^{-9}$ | 0.1616 |
| $CoFe_{1.8}Cr_{0.2}O_4$ | 323 | 9.778 | $5.45 \times 10^{-9}$ | 10372 | $2.25 \times 10^{-10}$ | $4.20 \times 10^6$ | $1.30 \times 10^{-8}$ | 0.309 |
| | 373 | 10.55 | $2.04 \times 10^{-8}$ | 9999 | $2.26 \times 10^{-10}$ | $2.51 \times 10^5$ | $1.93 \times 10^{-8}$ | 0.384 |
| | 423 | 11.98 | $3.81 \times 10^{-10}$ | 9346 | $5.77 \times 10^{-10}$ | $2.16 \times 10^4$ | $1.29 \times 10^{-6}$ | 0.1933 |
| | 473 | 12.77 | $2.74 \times 10^{-10}$ | 2820 | $1.84 \times 10^{-9}$ | 1338 | $4.16 \times 10^{-7}$ | 0.2669 |
| | 523 | 11.63 | $1.25 \times 10^{-9}$ | 25.41 | $2.04 \times 10^{-10}$ | 971.7 | $4.59 \times 10^{-5}$ | 0.6997 |

| | | | | | | | |
|---|---|---|---|---|---|---|---|
| | 573 | 18.43 | $1.95 \times 10^{-8}$ | 2.136 | $2.85 \times 10^{-10}$ | 437 | $2.30 \times 10^{-6}$ | 7.123 |
| | 323 | 14.07 | $1.65 \times 10^{-9}$ | $9.73 \times 10^{5}$ | $3.04 \times 10^{-10}$ | $1.73 \times 10^{6}$ | $2.78 \times 10^{-7}$ | 0.1739 |
| | 373 | 14.92 | $1.15 \times 10^{-9}$ | $2.938 \times 10^{4}$ | $3.44 \times 10^{-10}$ | $7.15 \times 10^{5}$ | --- | --- |
| $CoFe_{1.7}Cr_{0.3}O_4$ | 423 | 15.7 | $1.25 \times 10^{-9}$ | 4705 | $3.46 \times 10^{-10}$ | $9.66 \times 10^{4}$ | $6.44 \times 10^{-7}$ | 0.1038 |
| | 473 | 16.93 | $8.79 \times 10^{-10}$ | 1779 | $4.14 \times 10^{-10}$ | 6160 | $9.07 \times 10^{-5}$ | 0.002877 |
| | 523 | 17.54 | $1.99 \times 10^{-10}$ | 1293 | $2.96 \times 10^{-10}$ | 3346 | $4.70 \times 10^{-7}$ | 0.0005934 |
| | 573 | 13.11 | $2.89 \times 10^{-10}$ | 1069 | $7.39 \times 10^{-9}$ | 2138 | $2.10 \times 10^{-6}$ | $7.36 \times 10^{-11}$ |
| | 323 | 9.316 | $2.03 \times 10^{-9}$ | $1.192 \times 10^{6}$ | $3.04 \times 10^{-10}$ | $2.3 \times 10^{6}$ | $4.11 \times 10^{-7}$ | 0.1493 |
| | 373 | 14.9 | $4.33 \times 10^{-10}$ | $1.924 \times 10^{5}$ | $9.07 \times 10^{-10}$ | $2.64 \times 10^{5}$ | - | - |
| $CoFe_{1.6}Cr_{0.4}O_4$ | 423 | 15 | $1.32 \times 10^{-9}$ | $2.28 \times 10^{4}$ | $3.84 \times 10^{-10}$ | $1.25 \times 10^{5}$ | $1.96 \times 10^{-5}$ | 0.01866 |
| | 473 | 15.95 | $3.72 \times 10^{-10}$ | 5070 | $1.77 \times 10^{-9}$ | 2923 | $3.90 \times 10^{-6}$ | 0.003499 |
| | 523 | 16.09 | $3.25 \times 10^{-10}$ | 1205 | $2.01 \times 10^{-9}$ | 2623 | $8.82 \times 10^{-5}$ | 0.8349 |
| | 573 | 12.95 | $3.29 \times 10^{-10}$ | 23.92 | $5.44 \times 10^{-7}$ | 68.41 | $5.49 \times 10^{-6}$ | 0.3698 |

Table V. Activation energy of $CoFe_{2-x}Cr_xO_4$ for x= 0, 0.1, 0.2, 0.3 & 0.4 samples.

| Sample | Activation Energy in eV ($E_A$) taken at 533669 Hz from $\ln\sigma_{ac}$ vs 1000/T plot. | | | $Z''$ Vs. $\omega_{max}$ data | |
|---|---|---|---|---|---|
| | Region I (below 423 K) | Region II (423 K to 543K) | Region III (543K to 673 K) | $E_A$(from log $\omega_{max}$ Vs 1000/T plot) | |
| | | | | Region I (323 K- 523 K) | Region II (323 K- 523 K) |
| x=0.0 | 0.29 | 0.54 | 0.15 | 0.62 | 0.12 |
| x=0.1 | 0.34 | 0.50 | 0.28 | 0.57 | 0.32 |
| x=0.2 | 0.33 | 0.55 | 0.37 | 0.61 | 0.10 |
| x=0.3 | 0.26 | 0.48 | 0.61 | 0.46 | 0.24 |
| x=0.4 | 0.39 | 0.20 | 0.69 | 0.61 | 0.24 |

Table VI. The value of resistivity and Mott characteristic temperature for different doping concentration.

| Sample | $\rho_{300K}$($\Omega$cm) | Region I | | Region II | |
|---|---|---|---|---|---|
| | | $\rho_o$($\Omega$cm) | $T_0$(x$10^{10}$ K) | $\rho_o$($\Omega$cm) | $T_0$(x$10^9$ K) |
| x=0.0 | 39788 | 4.19x$10^{-32}$ | 1.47 | 3.75x$10^{-10}$ | 1.60 |
| x=0.1 | 42614 | 6.00x$10^{-32}$ | 1.49 | 2.09x$10^{-11}$ | 2.60 |
| x=0.2 | 161485 | 1.53x$10^{-33}$ | 1.62 | 2.12x$10^{-19}$ | 2.40 |
| x=0.3 | 2.23x$10^7$ | 4.50x$10^{-33}$ | 1.71 | 1.08x$10^{-25}$ | 7.30 |
| x=0.4 | 2.35x$10^7$ | 6.77x$10^{-33}$ | 1.85 | 2.79x$10^{-27}$ | 10.04 |

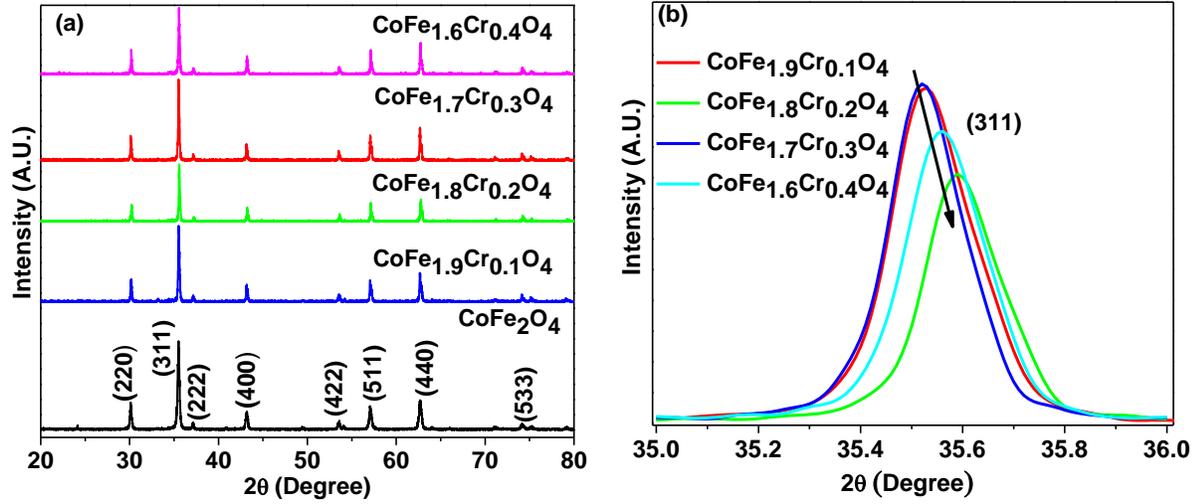

FIG. 1. (a) X-ray diffraction (XRD) patterns of $CoFe_{2-x}Cr_xO_4$ for x=0.0, 0.1, 0.2, 0.3 and 0.4 samples annealed at 1173 K and patterns are indexed to $Fd\bar{3}m$ spacegroup in cubic symmetry.

FIG. 1. (b) Highest intensity peak (311) in the XRD patterns of $CoFe_{2-x}Cr_xO_4$ for x=0.1, 0.2, 0.3, 0.4 samples.

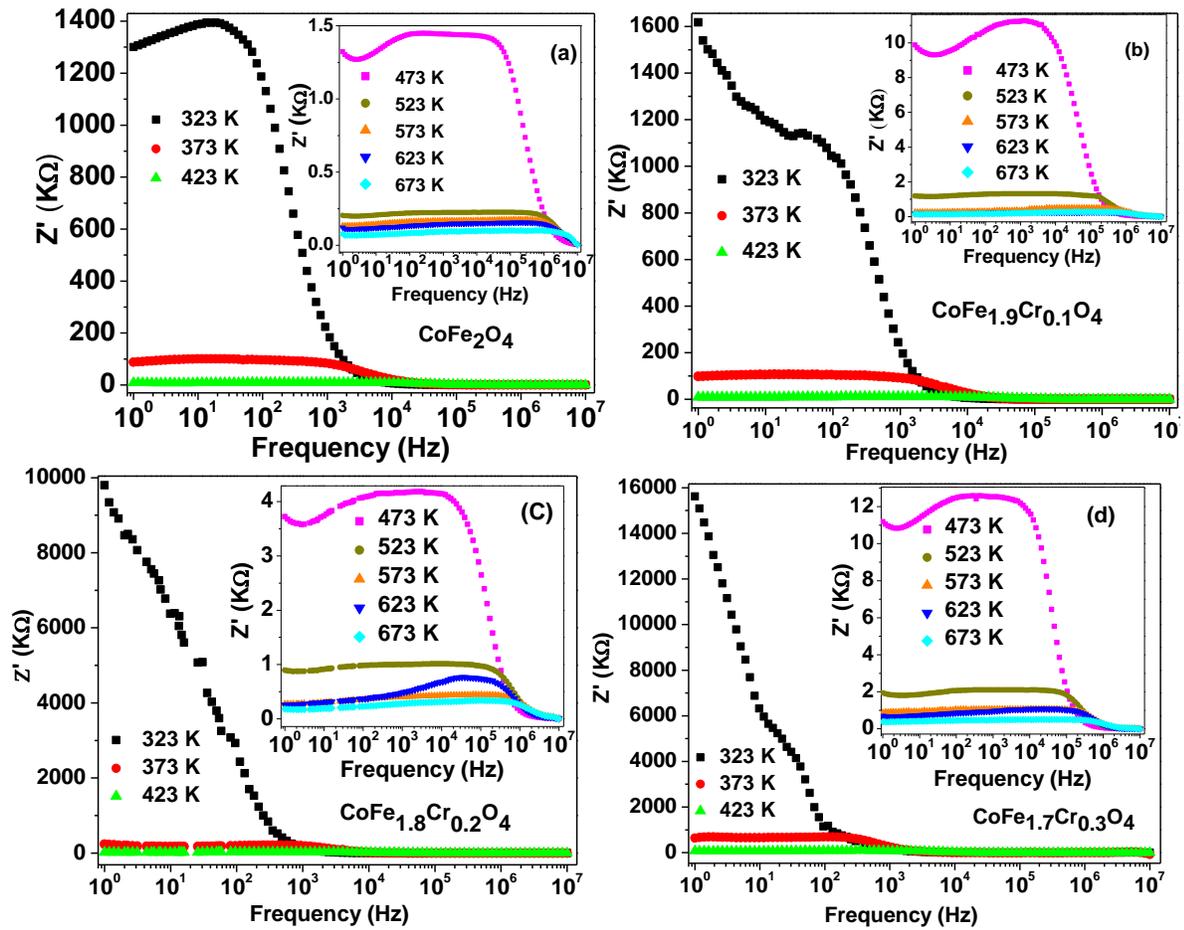

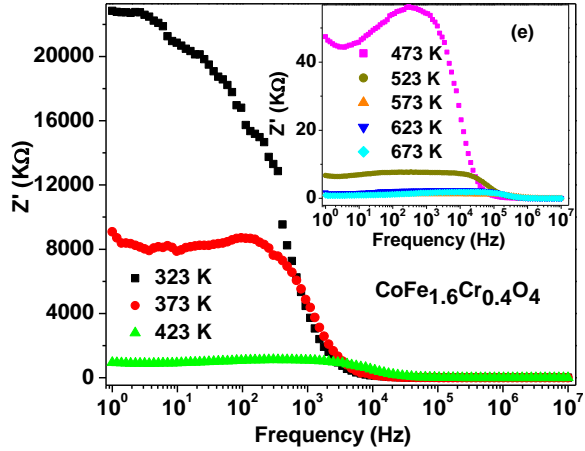

FIG. 2. Frequency variation of the real part ($Z'$) of Impedance spectrum of $CoFe_{2-x}Cr_xO_4$ for (a)x=0.0, (b)x=0.1, (c)x=0.2, (d)x=0.3 and (e)x=0.4 at different temperatures below 450 K. Insets depict the frequency variation of $Z'$ at different temperatures within the range 473 K to 673K.

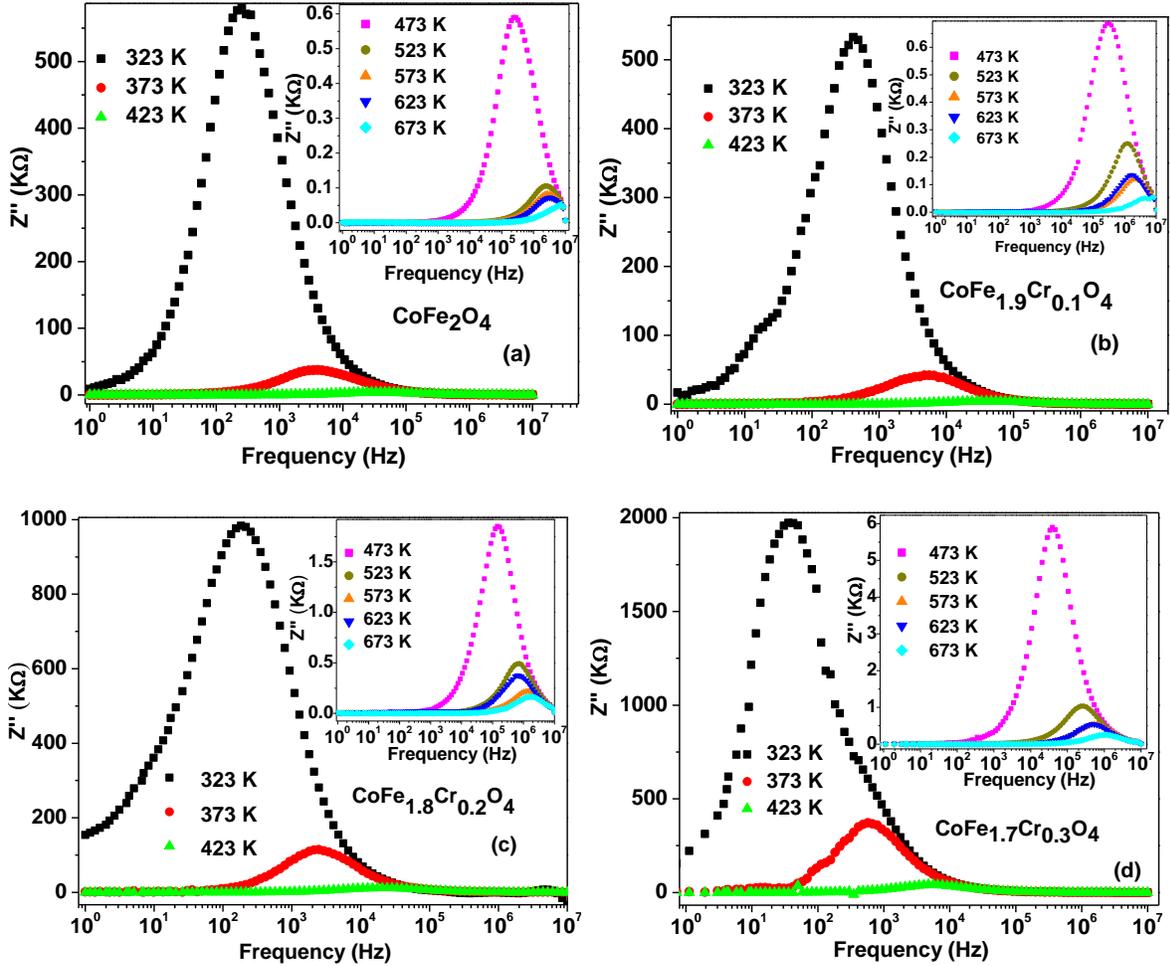

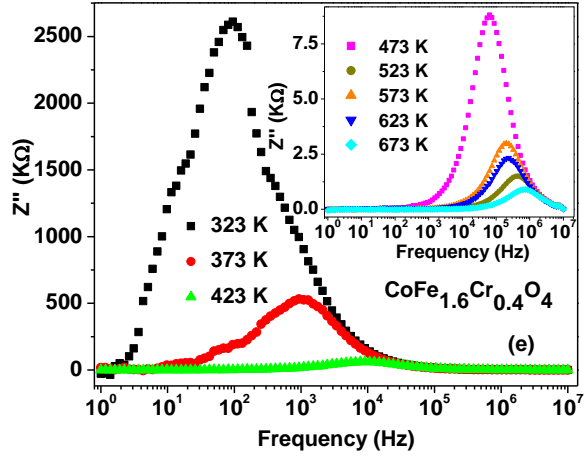

FIG. 3. Frequency variation of the imaginary part ($Z''$) of Impedance spectrum of $CoFe_{2-x}Cr_xO_4$ for (a)x=0.0, (b)x=0.1, (c)x=0.2, (d)x=0.3 and (e)x=0.4 at different temperatures. Insets show the frequency variation of $Z''$ from temperature 473 K to 673 K.

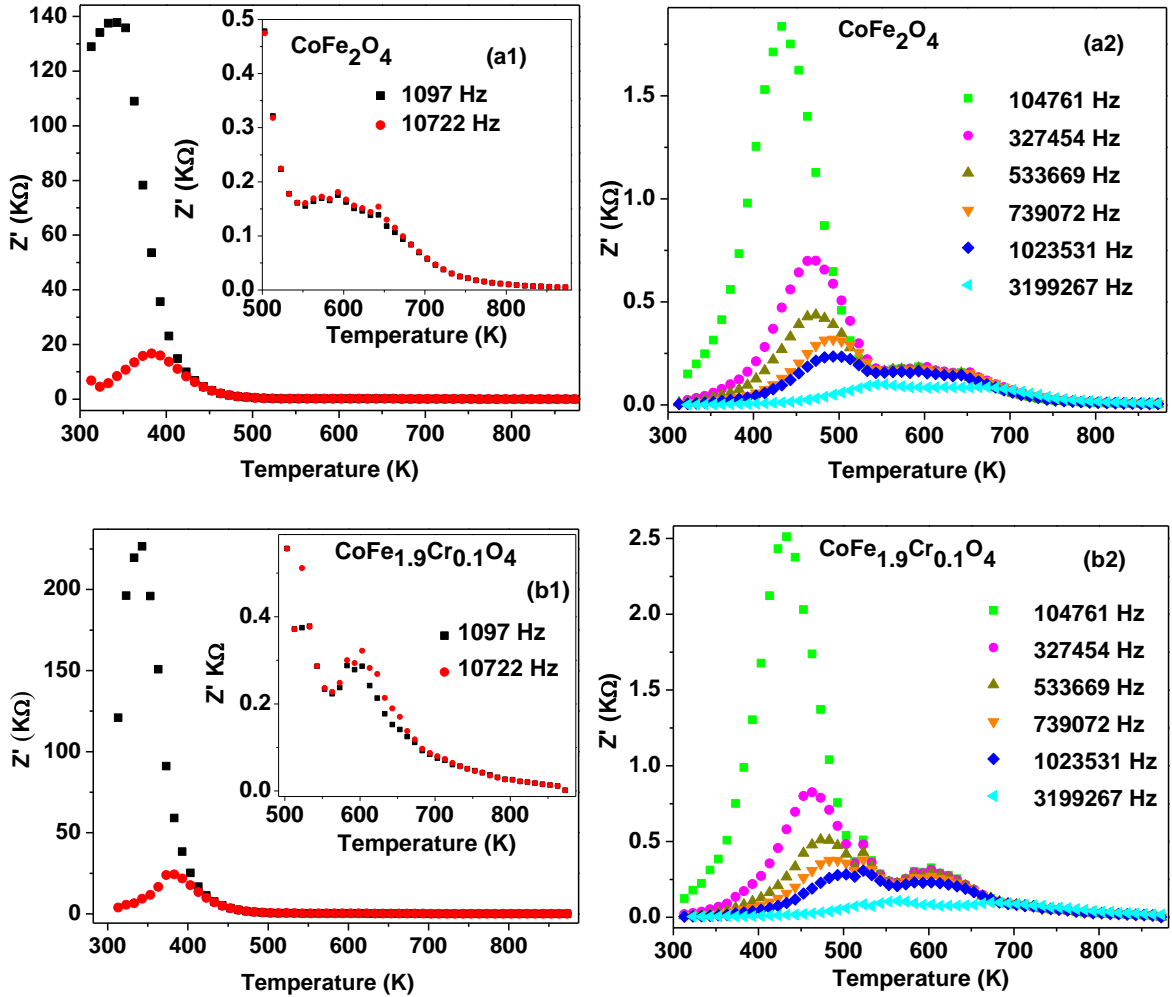

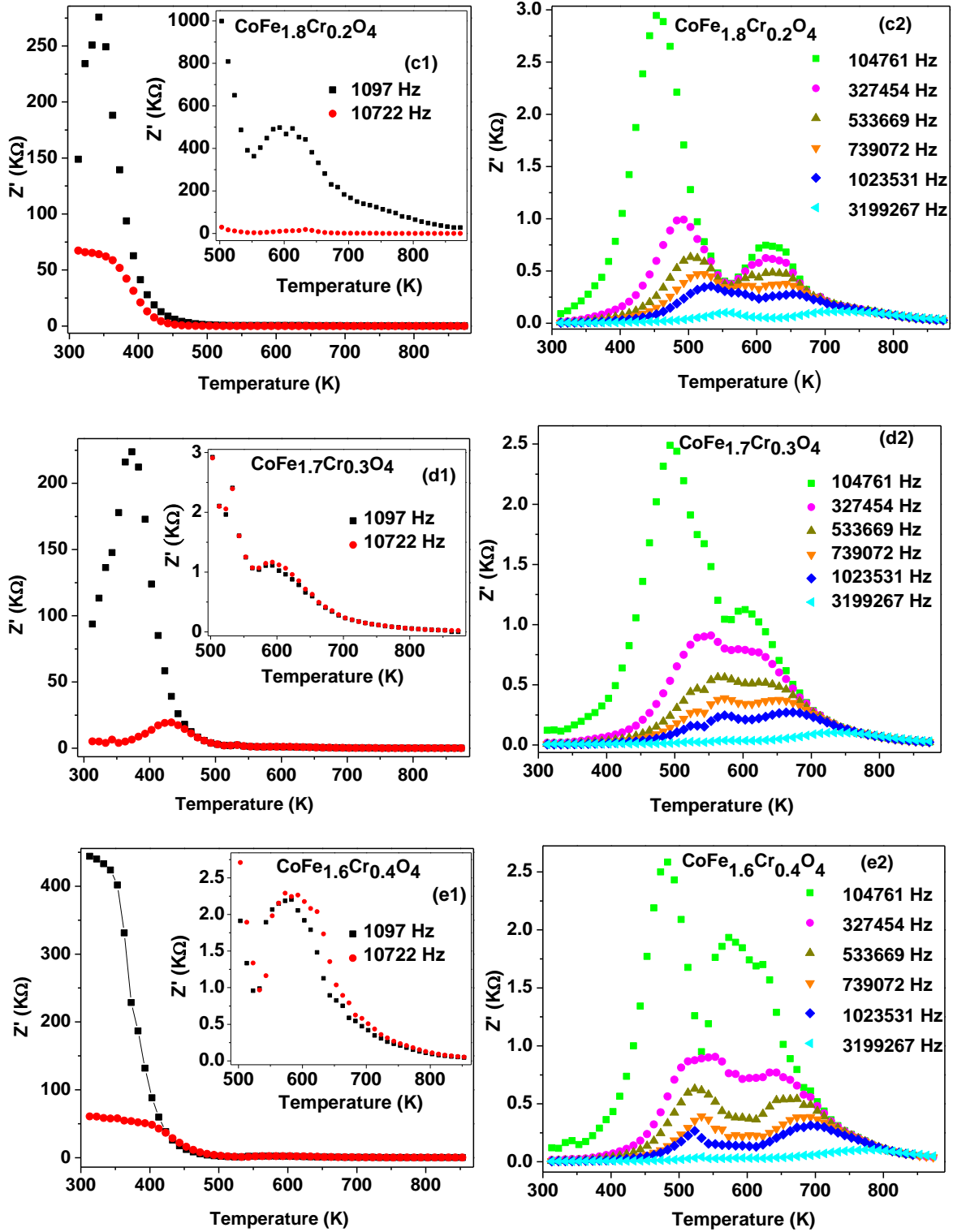

FIG. 4. Temperature dependent real part (Z′) of Impedance for $CoFe_{2-x}Cr_xO_4$ for (a)x=0.0, (b)x=0.1, (c)x=0.2, (d)x=0.3 and (e)x=0.4. Insets are the temperature dependent Z′ from 500-800 K to distinguish the second peak($T_M$).

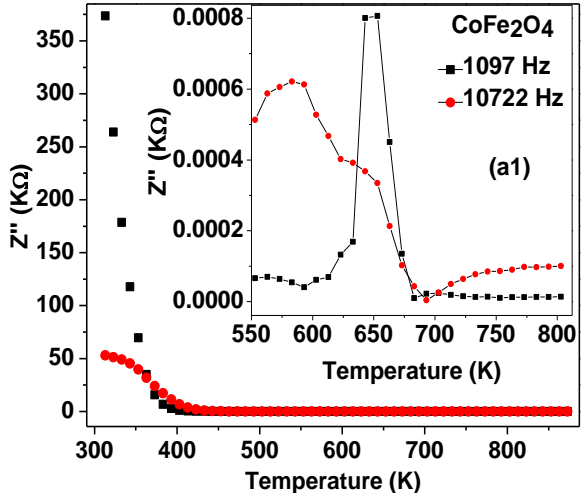
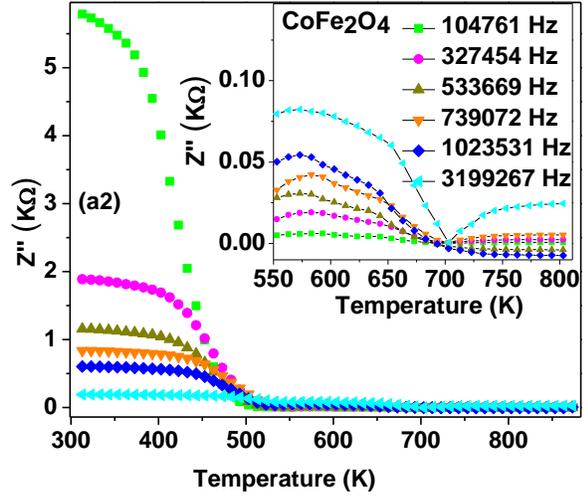
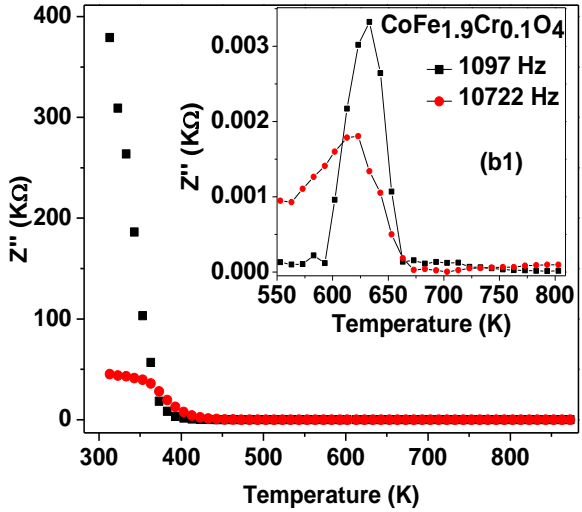
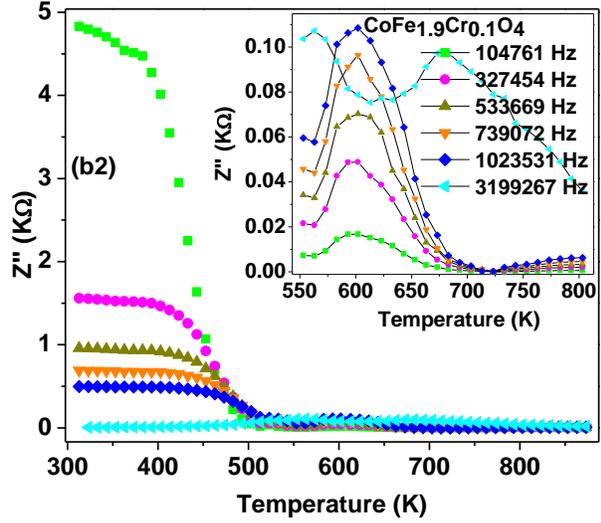
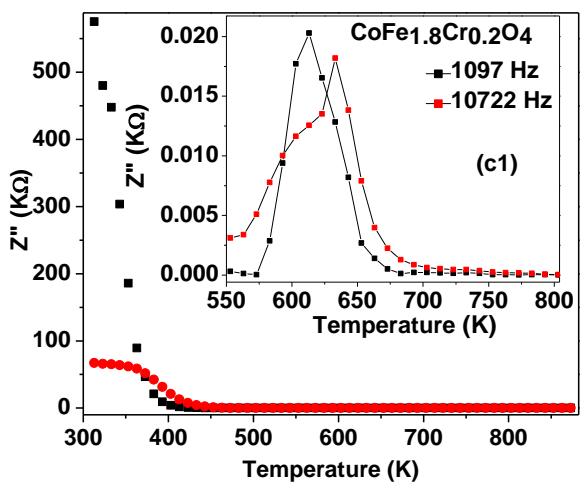
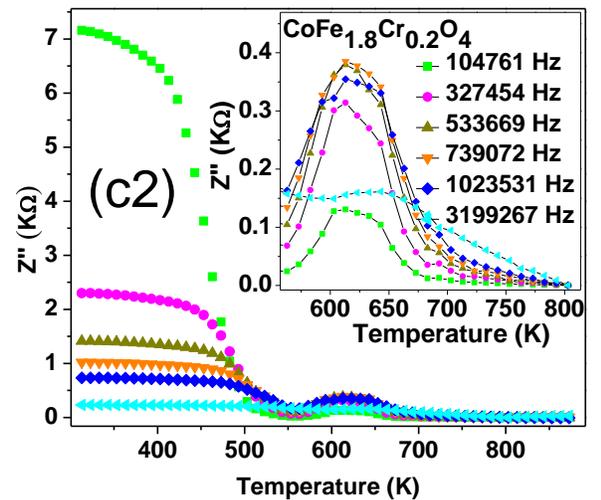

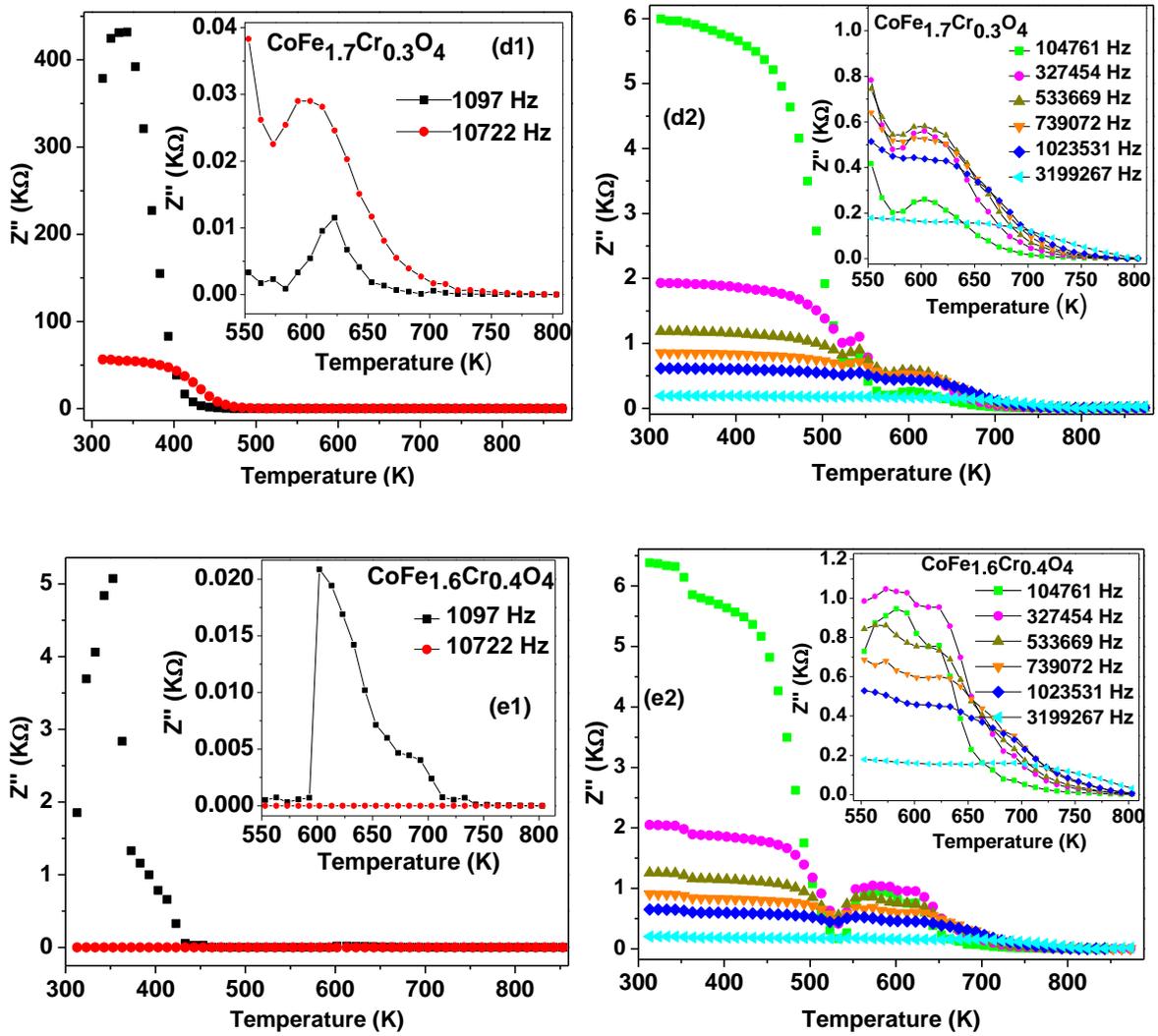

FIG. 5. Temperature dependent imaginary part ($Z''$) of impedance of $CoFe_{2-x}Cr_xO_4$ for (a)x=0.0, (b)x=0.1, (c)x=0.2, (d)x=0.3 and (e)x=0.4. Insets are the temperature dependent $Z''$ spectrum of $CoFe_{2-x}Cr_xO_4$ for x=0.0-0.4 at different frequencies.

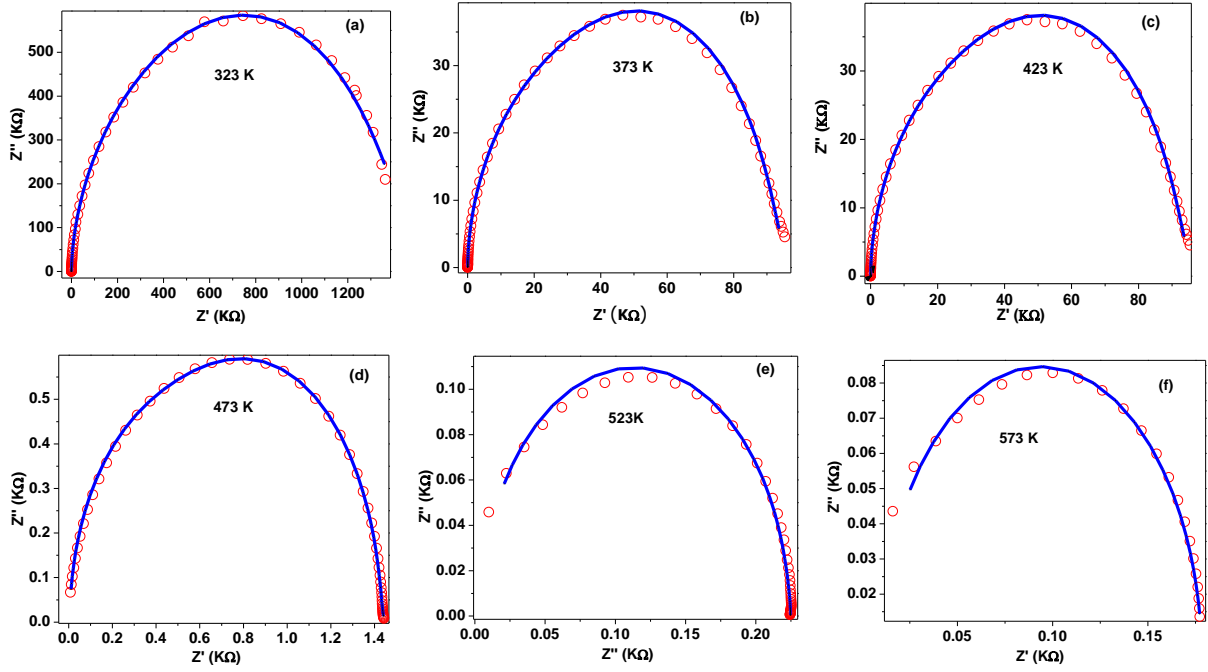

FIG. 6. (a) Cole-Cole plots or complex impedance spectra for sample $CoFe_2O_4$ at different temperatures.

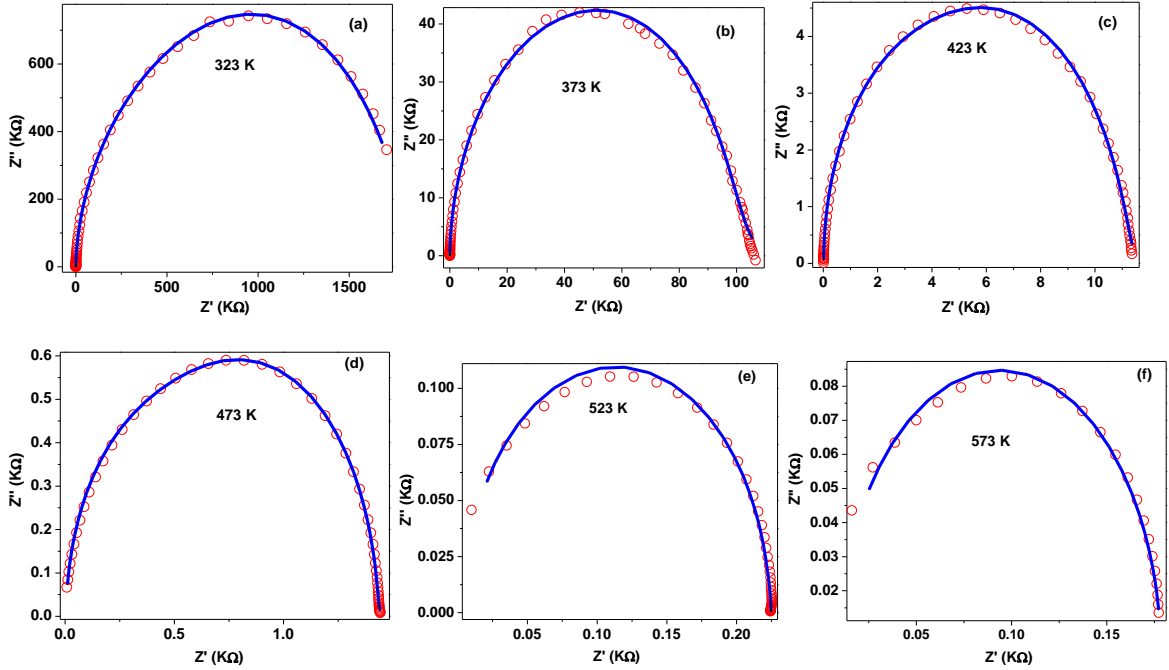

FIG. 6. (b) Cole-Cole plots or complex impedance spectra for sample $CoFe_{1.9}Cr_{0.1}O_4$ at different temperatures.

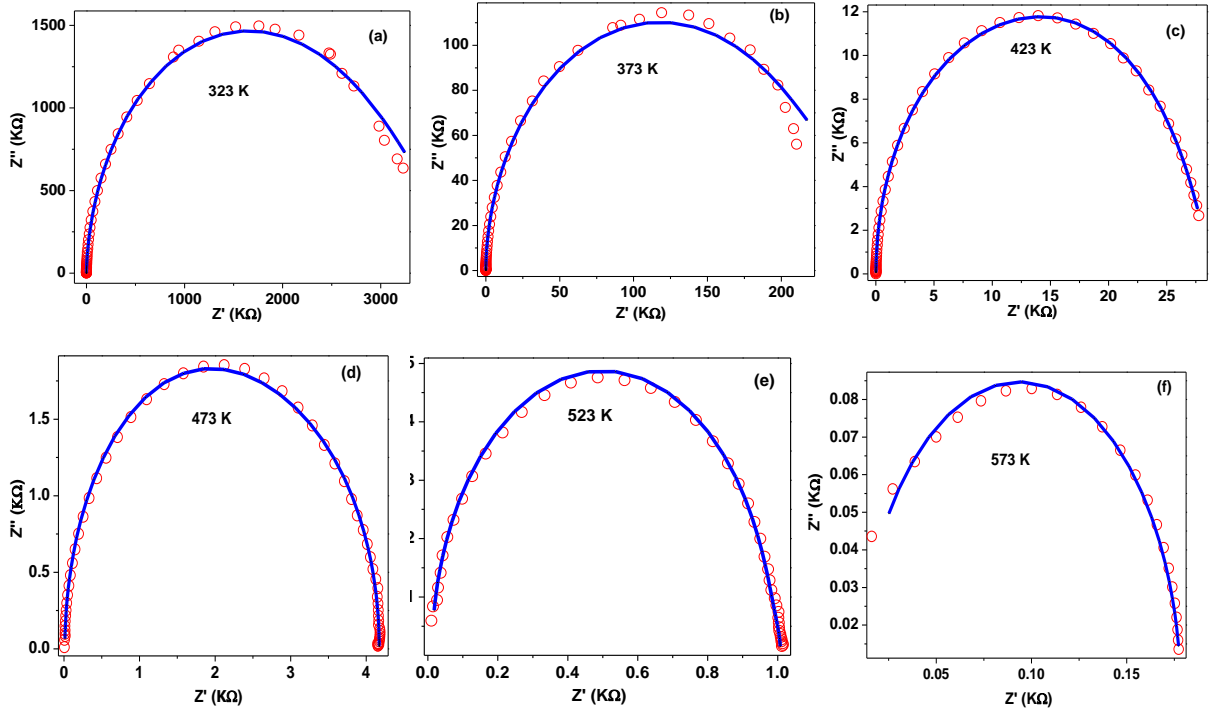

FIG. 6. (c) Cole-Cole plots or complex impedance spectra for sample $CoFe_{1.8}Cr_{0.2}O_4$ at different temperatures.

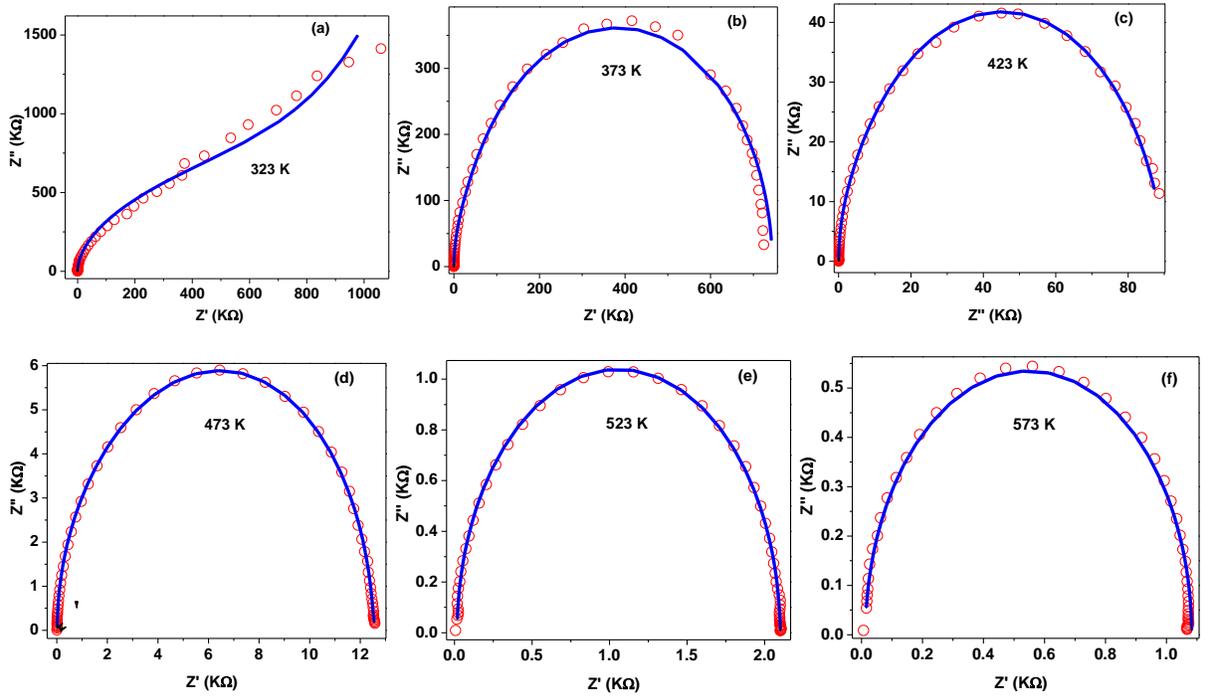

FIG. 6. (d) Cole-Cole plots or complex impedance spectra for sample $CoFe_{1.7}Cr_{0.3}O_4$ at different temperatures.

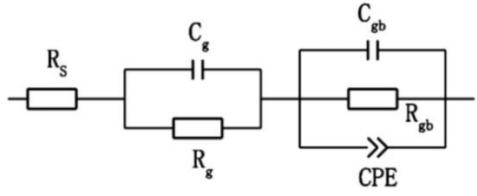

**FIG. 7.** Equivalent Circuit model.

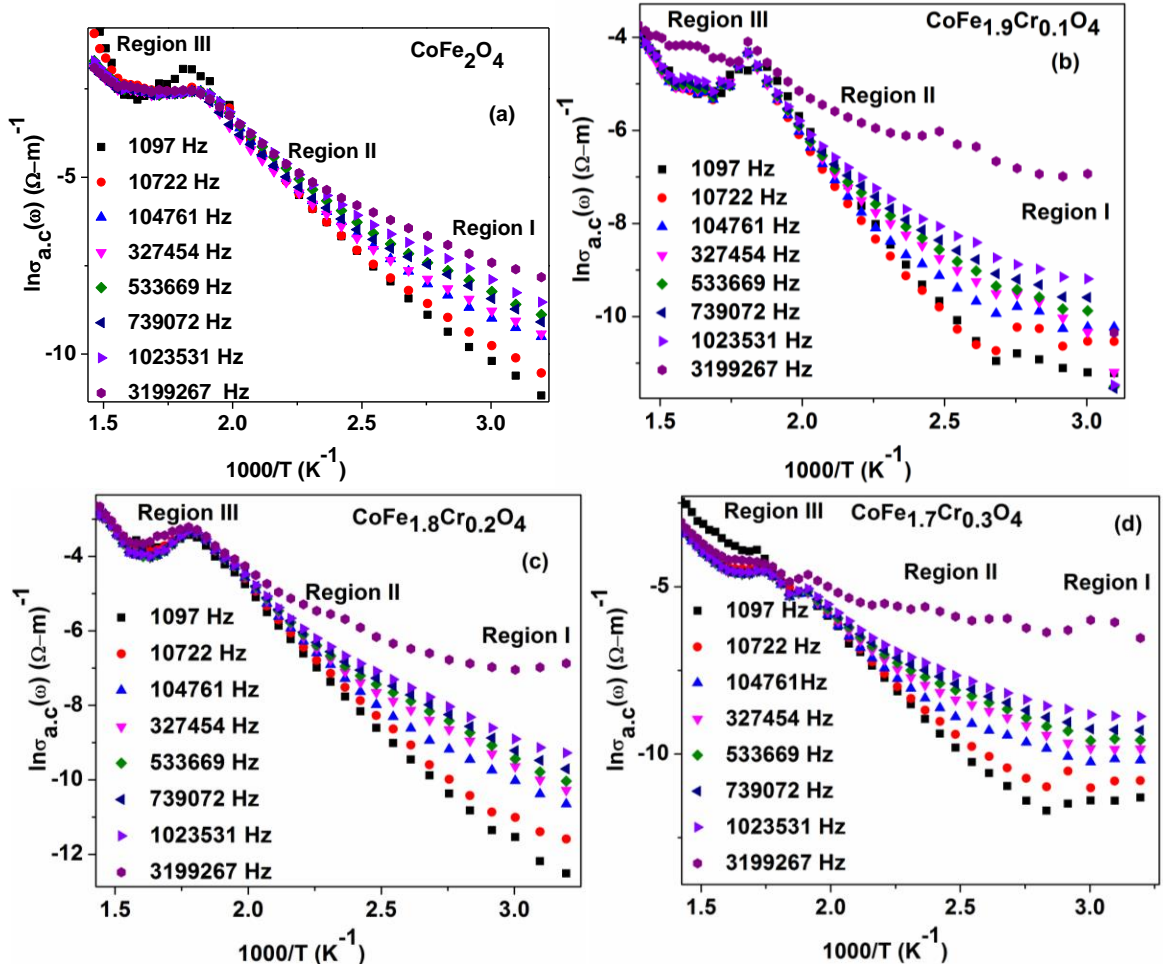

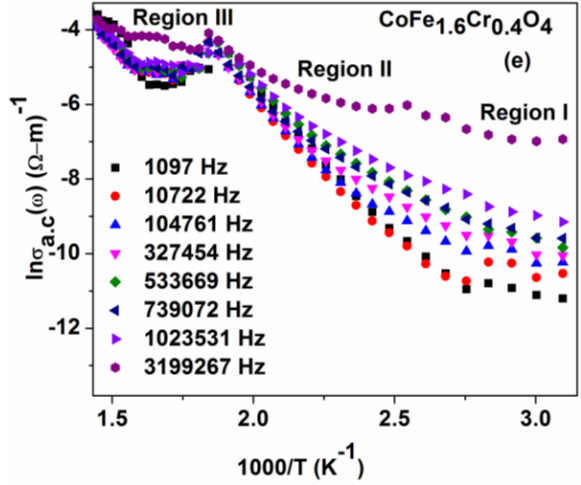

FIG. 8. Temperature dependent of ac conductivity ($\sigma_{ac}$) of CoFe$_{2-x}$Cr$_x$O$_4$ for (a) x=0.0, (b)x=0.1, (c)x=0.2, (d)x=0.3 and (e)x=0.4.

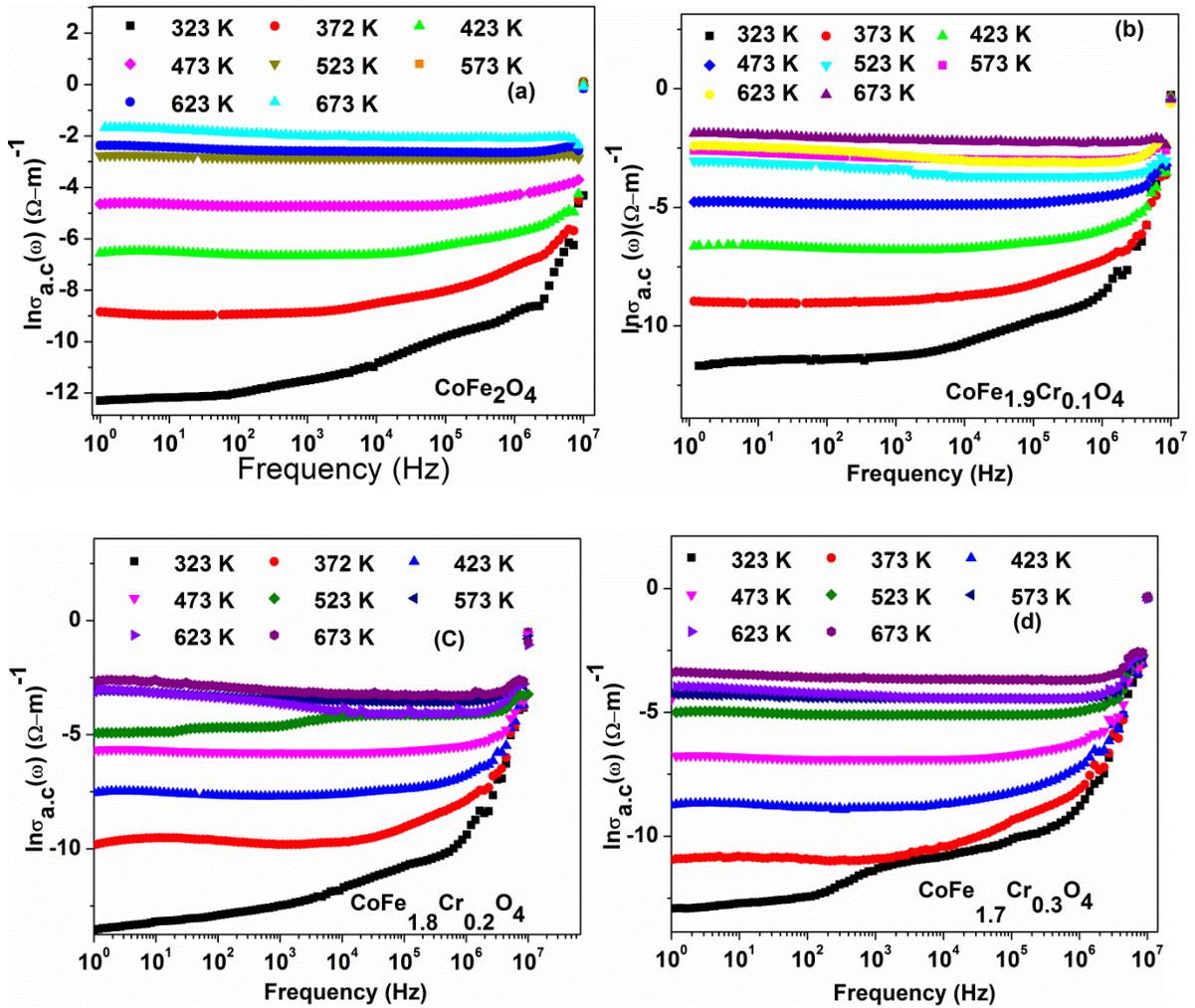

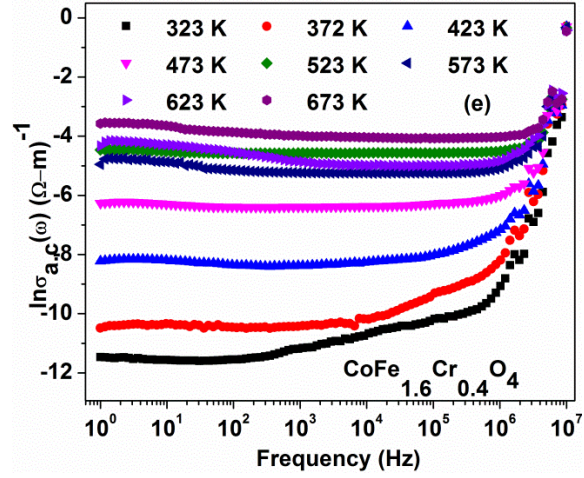

FIG. 9. Frequency dependent conductivity of $CoFe_{2-x}Cr_xO_4$ for (a) x=0.0, (b)x=0.1, (c)x=0.2, (d)x=0.3 and (e)x=0.4 at different temperatures.

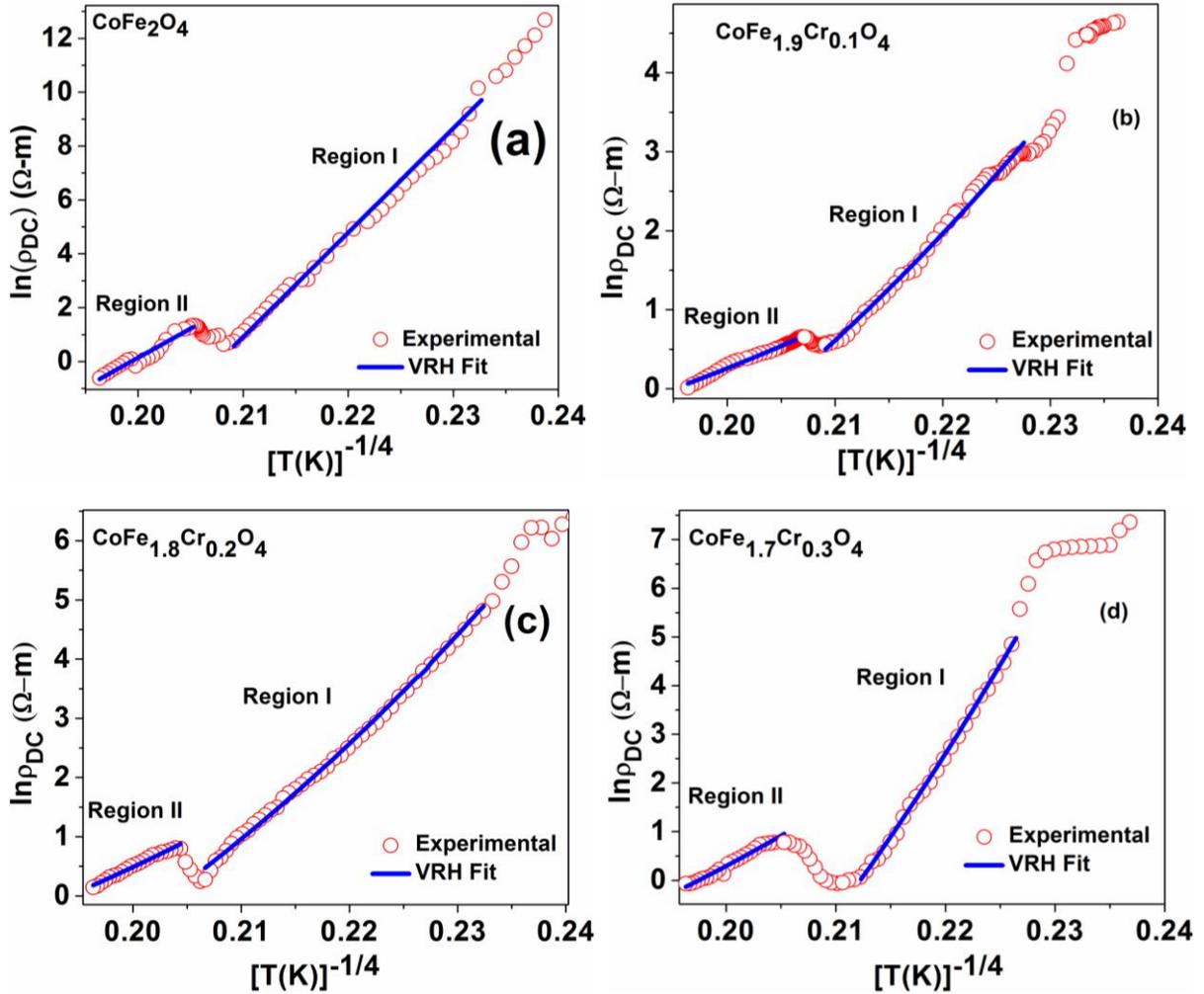

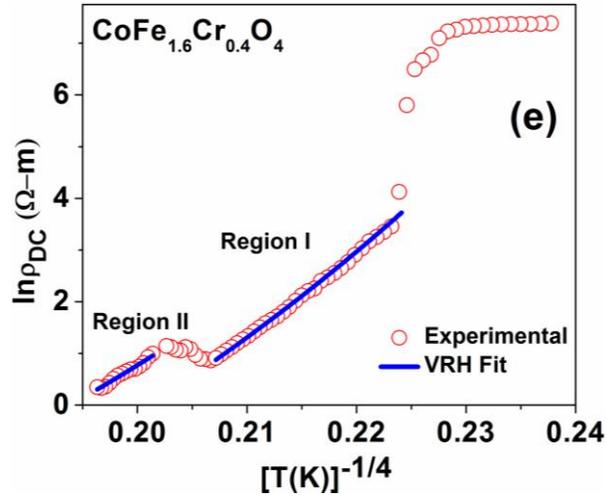

FIG. 10. Logarithmic DC resistivity (ln$\rho_{DC}$) versus Temperature $[T(K)]^{-1/4}$ plot of DC resistivity. Solid lines represent the analyzed data point for VRH model.